\documentclass[fleqn,usenatbib]{mnras}
\usepackage{newtxtext,newtxmath}
\usepackage[T1]{fontenc}
\DeclareRobustCommand{\VAN}[3]{#2}
\let\VANthebibliography\thebibliography
\def\thebibliography{\DeclareRobustCommand{\VAN}[3]{##3}\VANthebibliography}
\usepackage{graphicx}
\usepackage{amsmath}
\title[Two cyclotron absorption lines in Cen X-3]{Discovery of Two Cyclotron Resonance Scattering Features in X-ray Pulsar Cen X-3 by Insight-HXMT
}
\author[W. Yang et al.]{W. Yang$^{1,2}$, W. Wang$^{1,2}$\thanks{E-mail: wangwei2017@whu.edu.cn}, Q. Liu$^{1,2}$, X. Chen$^{1,2}$, H. J. Wu$^{1,2}$, P. F. Tian$^{1,2}$, J. S. Chen$^{1,2}$ \\
$^{1}$Department of Astronomy, School of Physics and Technology, Wuhan University, Wuhan 430072, China \\
$^{2}$WHU-NAOC Joint Center for Astronomy, Wuhan University, Wuhan, 430072, China}
\date{Accepted XXX. Received YYY; in original form ZZZ}
\pubyear{2022}
\begin{document}
\begin{sloppypar}
\label{firstpage}
\pagerange{\pageref{firstpage}--\pageref{lastpage}}
\maketitle

\begin{abstract}
We present the results of the neutron star X-ray binary system Cen X-3 performed by $Insight$-HXMT with two observations during 2017 and 2018. During these two observations, the source reached a X-ray luminosity of $\sim 10^{38}$ erg s$^{-1}$ from 2 -- 105 keV. The analysis of the broadband X-ray spectrum reports the presence of two cyclotron resonance scattering features (CRSFs) with the fundamental line at $\sim$ 28 keV and the harmonic line at $\sim 47 $ keV. The multiple lines exist by fittings with different continuum models, like the absorbed NPEX model and a power-law with high energy exponential cutoff model. This is the first time that both fundamental and harmonic lines are detected in Cen X-3. We also show evidence of two cyclotron lines in the phase-resolved spectrum of Cen X-3. The CRSF and continuum spectral parameters show evolution with the pulse profile, and the two line centroid energy ratio does not change significantly and locates in a narrow value range of $1.6-1.7$ over the pulse phase. The implications of the discovering two cyclotron absorption features and phase-resolved spectral properties are discussed.
\end{abstract}


\section{Introduction}
The accreting X-ray pulsars have strong magnetic field, and when the magnetic pressure dominates the gas in the accretion disk, the flow of infalling gas is channeled along magnetic field lines to the surface of neutron star. Most of the gas fall on an area which is less than the whole surface area of the neutron and form the "hot spot" where the kinetic energy of gas is converted into radiation. Due to the magnetic axis and rotation axis of the neutron star may not be parallel, as the star rotates, the radiation from the hot spots forms a pulse profile with the same rotation period of neutron star \citep{davidson1973accretion}. The Larmor radius of infalling electrons becomes comparable to or smaller than the electrons' de Broglie wavelength in such a strong magnetic field, then the quantum effects of electronic orbital radius and energy become unignorable \citep{1992High}. The photons are resonantly scattered by the electrons near the neutron star surface and absorption lines will appear on the X-ray spectra, the so-called cyclotron resonance scattering features (CRSFs). The energy of fundamental lines which correspond to the cyclotron scattering from the ground level to the first excited Landau level is relevant to the magnetic field as
\begin{flalign}
&\ E_{cyc}\simeq 11.6\ {\rm keV}\times B_{12}\left( 1+z \right) ^{-1}&
\end{flalign}
where $B_{12}$ is the magnetic field in units of $10^{12}\ \rm G$, and $z$ is the gravitational redshift near the surface of a neutron star.
\par
The neutron star system Cen X-3 discovered by Uhuru \citep{giacconi1971discovery} has a spin period of $\sim 4.8\ \rm s$ \citep{1972Evidence} and an orbital period of $\sim 2.08\ \rm d$ \citep{Falanga2015EphemerisOD}, is a high-mass X-ray binary pulsar located at a distance of $\sim 6.9$ kpc based on the recent Gaia measurement \citep{van2021new}. The system consists of a neutron star with the mass of 1.21±0.21 $M_{\odot}$ and an O-type supergiant v779 Cen of mass 20.5±0.7 $M_{\odot}$ \citep{ash1999mass} with radius of 11.4$\pm$0.7$R_{\odot}$ \citep{Falanga2015EphemerisOD}. The eclipse lasts for 20$\%$ of the nearly circular orbit \citep{nagase1989accretion} with the orbital eccentricity below 0.0016 \citep{Bildsten_1997}. The mass transfer between the high luminosity of Cen X-3 and supergiant via Roche lobe overflow is inferred from the optical light curves of this system \citep{tjemkes1986optical}. The pulse profile shape also changes with energy, the pulse profile is clearly single-peaked at high energy (roughly above 10 keV), and double peak at low energy observed by $Ginga$ \citep{nagase1992ginga} and $BeppoSAX$ \citep{Burderi_2000}. Cen X-3 is a bright X-ray source, with the observed X-ray average luminosity about $(0.1-1)\times10^{38}$ erg s$^{-1}$ \citep{white1983accretion} and can sometimes locate in the low luminosity \citep{nagase1992ginga}. 
\par
The cyclotron absorption line near 30 keV has been observed in the spectrum of Cen X-3 through several X-ray missions. \cite{nagase1992ginga} reported the presence of an absorption line at $\sim$ 30 keV using the combination of the power law continuum and a Lorenzian shape by Ginga, which later was identified as cyclotron resonant scattering by BeppoSAX \citep{santangelo1998bepposax}. RXTE observations \citep{Suchy_2008} also suggested a cyclotron resonance scattering feature at about 31 keV with a continuum described by the Fermi-Dirac cut-off model ($fdcut$, \citealt{tanaka_1986}). \cite{10.1093/mnras/staa3477} used the spectra of Suzaku and NuSTAR, and determined the cyclotron line energy $\sim$ 30.29 keV and 29.22 keV, respectively. This is also acknowledged that the spectrum of an X-ray pulsar varies with the pulse phase. \cite{Burderi_2000} observed the cyclotron line energy decreasing from 36 to 28 keV along the peak of pulse profile. During the RXTE observations by \cite{Suchy_2008}, the cyclotron line energy of Cen X-3 increased during the early rise of the main peak, followed by a gradual decrease. The CRSF energy of Cen X-3 does not show a tendency of decrease/increase during the ten years of observations \citep{ji2019long}. It is still an open question whether there are multiple cyclotron features in Cen X-3, and there is no evidence of other absorption features in the spectrum for Cen X-3 in previous observations yet.
\par
In this paper we report the results of a spectral analysis of the broad band spectrum ($2-105$ keV) of Cen X-3 observed with the Insight-HXMT. We confirm the presence of two cyclotron absorption features at $\sim 28$ and $\sim 47$ keV in the spectrum of Cen X-3 at the first time. In Section 2 we present the observations and the data extraction. Timing analysis and the X-ray spectral analysis including phase-averaged spectrum and phase-resolved spectroscopy are presented in section 3. We make a conclusion of the discovery of two cyclotron lines in Cen X-3 and a brief discussion is shown in Section 4.

\section{OBSERVATIONS}
Insight-HXMT is the first X-ray astronomical satellite in China, launched on 2017 June 15. Three kinds of main scientific payloads aboard the Insight-HXMT consist of the High Energy X-ray telescope (HE) operating in $20-250$ keV \cite{liu2020high}, the Medium Energy X-ray telescope (ME) operating in $5-30$ keV \cite{cao2020medium} and the Low Energy X-ray telescope (LE) operating in $1-15$ keV \cite{chen2020low}. The detector areas of the three telescopes are 5100, 952, and 384 $cm^{2}$, respectively. HE detectors have the large area for hard X-ray studies, with the energy resolution of $\sim$15$\%$ at 60 keV.
\par
For each observation, the Insight-HXMT Data Analysis Software (HXMTDAS) v2.04 is used to analyze data (more details on the analysis processes can refer to the previous work, e.g., \citealt{WANG20211}; \citealt{chen2021relation}). In order to take advantage of the best screened event file to generate the high-level products including the energy spectra, response file, light curves and background files, we use tasks $he/me/lepical$ to remove spike events caused by electronic systems and $he/me/legtigen$ be utilized to select good time interval (GTI) when the pointing offset angle $< 0.04^\circ$; the pointing direction above earth $> 10^\circ$; the geomagnetic cut-off rigidity >8 GeV and the South Atlantic Anomaly (SAA) did not occur within 300 seconds. 
\par
Cen X-3 has been observed frequently by the pointing observations of Insight-HXMT. We concentrated on two observations identified as the Exposure IDs P010131100111 on July 26, 2017 and P010131101602 on June 27, 2018 when Cen X-3 reach a mean X-ray luminosity of $\sim10^{38}$ ergs s$^{-1}$ from $2-105$ keV (also see the spectral analysis results in Section \ref{sec:Spectral analysis}). The ExpID P010131100111 lasted for $\sim$ 25 ks from MJD 57960 13:49:17 to 20:43:47 with the mean HE count rate of $\sim 251.4$ cts/sec, and the ExpID P010131101602 lasted 6.7 ks from MJD 58296 05:41:09 to 07:34:09 with the HE count rate of $\sim 199.9$ cts/sec. Since the two observations have high flux levels, the high S/N ratio will increase sensitivity and spectral line studies, particularly for HE detectors, which makes it possible for us to measure the hard X-ray energy spectrum of the source with high accuracy. The final data analysis was performed with the XSPEC 12.11.1 version \citep{arnaud1996xspec}. All uncertainties are quoted at the 3$\sigma$ confidence level.

 
\begin{figure}
    \centering
    \includegraphics[width=.5\textwidth]{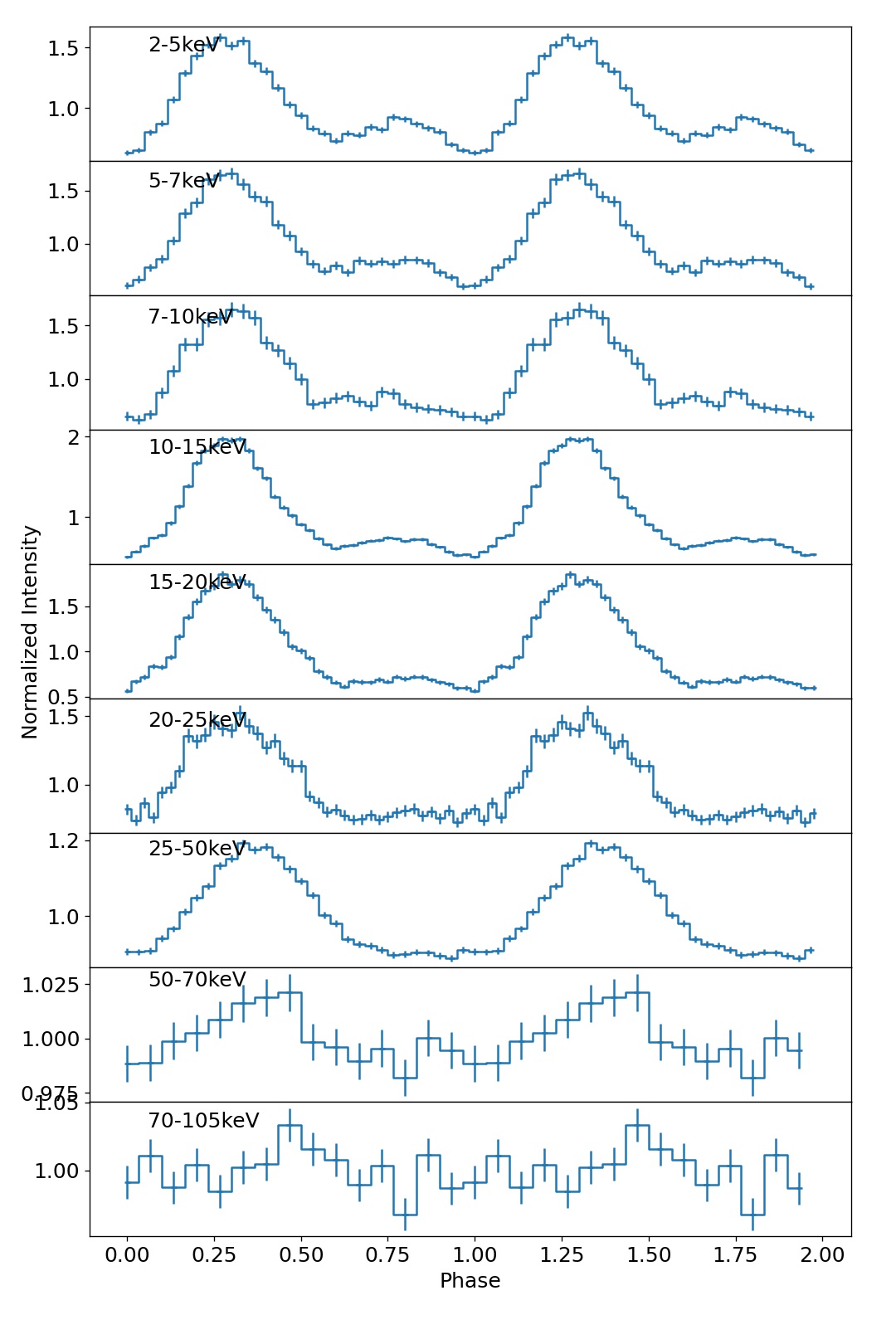}
    \caption{The pulse profiles of Cen X-3 for the ExpID: P010131101602 (MJD 58296) in different energy ranges.}
    \label{fig:1}
\end{figure}
\begin{figure*}
    \centering
    \includegraphics[width=0.48\textwidth,height=7.5cm]{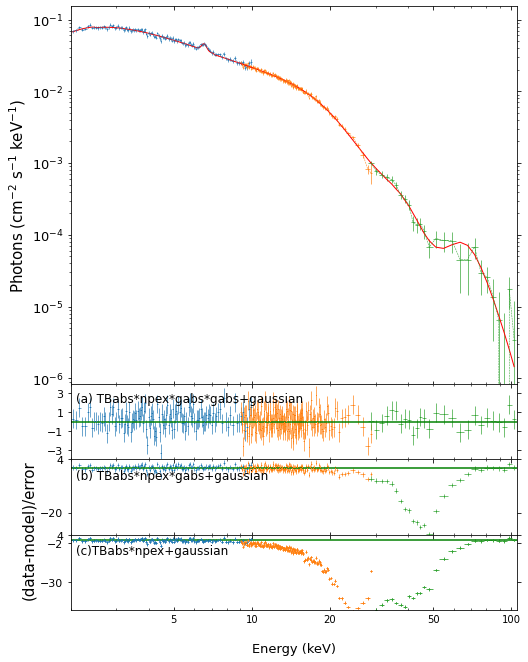}
    \includegraphics[width=0.48\textwidth,height=7.5cm]{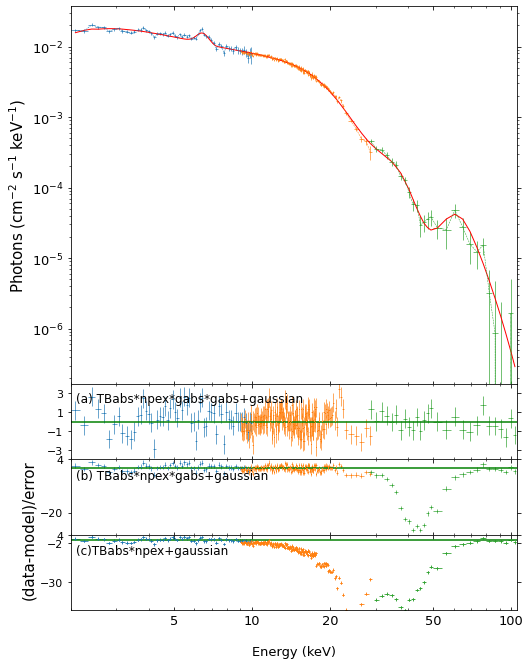}
    \includegraphics[width=0.48\textwidth,height=7.5cm]{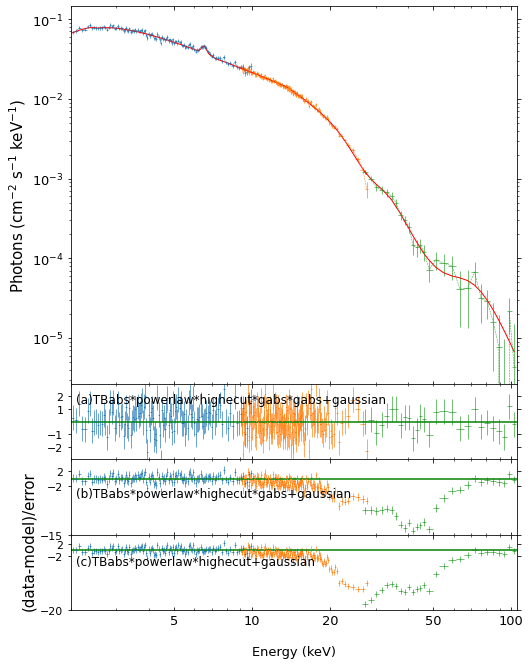}
    \includegraphics[width=0.48\textwidth,height=7.5cm]{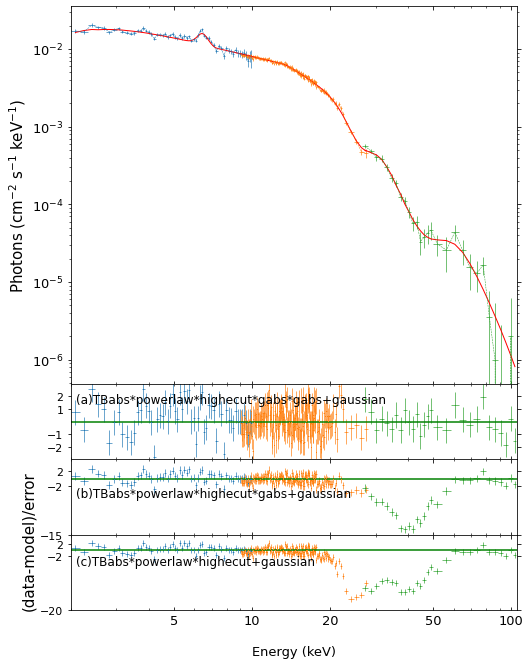}
    \caption{Fitted spectra of Cen X-3 with the energy range from 2 to 105 keV. The panels presented the spectra fitted by HIGHECUT (upper panel) and NPEX (bottom panel) models for two different observations with ExpID:  P010131101602(left); P010131100111(right). We also show residuals, i.e., the difference of data minus folded models divided by errors, the folded model with two cyclotron absorption features at $\sim 28$ keV and $\sim 47$ keV in panel (a), with only one cyclotron line at $\sim 28$ keV (b) and without the CRSFs (c), respectively.}
    \label{fig:2}
\end{figure*}
\begin{table*}
\caption{Best-fitting Parameters of Cen X-3 Observed with Insight-HXMT}
\label{table:1}
\begin{tabular}{cllllll}
\hline
\multicolumn{1}{l}{}                              & \multicolumn{2}{c}{HIGHECUT(gabs)}                                    & \multicolumn{2}{c}{NPEX(gabs)}                                        & \multicolumn{2}{c}{NPEX(cyclabs)}                                     \\ \cline{2-7} 
\multicolumn{1}{l}{}                              & \multicolumn{1}{c}{P010131101602} & \multicolumn{1}{c}{P010131100111} & \multicolumn{1}{c}{P010131101602} & \multicolumn{1}{c}{P010131100111} & \multicolumn{1}{c}{P010131101602} & \multicolumn{1}{c}{P010131100111} \\ \hline
\multicolumn{1}{l}{$N_{\rm H}$ ($10^{22}$ cm$^{-2}$)} & $2.31_{-0.13}^{+0.13}$            & $1.63_{-0.34}^{+0.15}$            & $2.06_{-0.12}^{+0.26}$            & $1.88_{-0.28}^{+0.38}$            & $2.09_{-0.30}^{+0.18}$            & $1.91_{-0.39}^{+0.25}$            \\
E$_{cyc1}$ (keV)                                  & $27.39_{-0.32}^{+2.67}$           & $26.29_{-0.83}^{+0.78}$           & $28.95_{-1.65}^{+2.82}$           & $28.19_{-1.25}^{+2.13}$           & $24.63_{-0.49}^{+0.78}$           & $24.83_{-0.34}^{+0.57}$           \\
$\sigma_{cyc1}$ (keV)                             & $2.86_{-1.56}^{+1.99}$            & $3.02_{-0.91}^{+0.48}$            & $7.63_{-0.88}^{+1.50}$            & $6.63_{-0.74}^{+1.31}$            & $14.98_{-3.26}^{+1.39}$           & $11.29_{-1.63}^{+2.00}$           \\
$\tau_1$                                          & $0.32_{-0.07}^{+0.36}$            & $0.70_{-0.22}^{+0.26}$            & $1.42_{-0.45}^{+0.83}$            & $1.49_{-0.40}^{+0.69}$            & $1.59_{-0.25}^{+0.04}$            & $1.61_{-0.14}^{+0.13}$            \\
E$_{cyc2}$ (keV)                                  & $46.90_{-3.47}^{+6.97}$           & $44.08_{-1.75}^{+1.24}$           & $48.73_{-2.98}^{+2.52}$           & $46.67_{-1.23}^{+2.44}$           & $46.54_{-2.58}^{+1.44}$           & $44.85_{-1.81}^{+1.88}$           \\
$\sigma_{cyc2}$ (keV)                             & $15.06_{-1.25}^{+3.87}$           & $8.21_{-1.70}^{+1.58}$            & $9.66_{-2.59}^{+1.40}$            & $8.08_{-1.34}^{+1.34}$            & $5.30_{-2.78}^{+4.63}$            & $7.01_{-2.62}^{+3.04}$            \\
$\tau_2$                                          & $1.45_{-0.34}^{+0.65}$            & $1.46_{-0.43}^{+0.32}$            & $2.30_{-0.66}^{+0.46}$            & $2.50_{-0.29}^{+0.50}$            & $1.58_{-0.32}^{+1.15}$            & $2.03_{-0.49}^{+0.28}$            \\
E$_{Fe K\alpha}$ (keV)                            & $6.54_{-0.07}^{+0.05}$            & $6.46_{-0.11}^{+0.04}$            & $6.54_{-0.06}^{+0.04}$            & $6.45_{-0.10}^{+0.05}$            & $6.54_{-0.08}^{+0.01}$            & $6.45_{-0.05}^{+0.04}$            \\
EqW$_{Fe K\alpha}$ (keV)                          & $0.15_{-0.08}^{+0.15}$            & $0.30_{-0.06}^{+0.20}$            & $0.15_{-0.06}^{+0.04}$            & $0.34_{-0.11}^{+0.15}$            & $0.15_{-0.08}^{+0.06}$            & $0.33_{-0.06}^{+0.08}$            \\
$\Gamma$                                          & $1.29_{-0.05}^{+0.02}$            & $0.86_{-0.04}^{+0.02}$            & $0.80_{-0.03}^{+0.08}$            & $0.61_{-0.11}^{+0.10}$            & $0.84_{-0.11}^{+0.03}$            & $0.66_{-0.13}^{+0.06}$            \\
E$_{c}$ (keV)                                   & $14.24_{-0.60}^{+0.78}$           & $13.46_{-0.28}^{+0.17}$           &     -      &    -      &     -       &    -    \\
E$_{f}$ (keV)                                  & $17.16_{-2.25}^{+1.64}$           & $12.42_{-1.18}^{+0.91}$           & $6.91_{-0.25}^{+0.21}$           & $6.49_{-0.14}^{+0.23}$     & $7.71_{-0.66}^{+0.17}$ & $7.18_{-0.30}^{+0.21}$         \\
Reduced-$\chi^2$ (dof)                            & 0.957(673)                        & 0.932(673)                        & 0.955(681)                        & 0.976(681)                        & 0.960(680)                        & 0.974(680)                        \\
\multicolumn{1}{l}{Flux (erg s$^{-1}$ cm$^{-2}$)}   & $1.04\times 10^{-8}$              & $3.43\times 10^{-9}$              & $1.39\times 10^{-8}$              & $5.34\times 10^{-9}$              & $1.85\times 10^{-8}$              & $7.02\times 10^{-9}$              \\ \hline
\end{tabular}
\end{table*}

\section{RESULTS}
\label{sec:Spectral analysis}
In this section we present the timing analysis and detailed spectral results including the phase-averaged and phase-resolved spectroscopy based on $Insight$-HXMT data, which aims to confirm two CRSFs discovered in Cen X-3.
\subsection{Energy-resolved pulse profile}
X-ray light curves for nine energy bands were extracted for the both ExpIDs. At first, we use HXMTDAS task $hxbary$ to change the photon arrival time from TT (Terrestrial Time) to TDB (Barycentric Dynamic Time) which considers the time delay due to the movement of the earth and satellite. And the orbital ephemeris given by \cite{Bildsten_1997} was adopted for the binary star correction. We search for the pulsed period of corrected light curves by the $efsearch$ (a built-in function in HEAsoft). The uncertainties of the spin period are estimated using a Gaussian error. Therefore, the pulse periods for the different bands of the phase-resolved spectroscopy are determined. We found that the source pulsates with a spin period of $4.797\pm 0.003$s for ExpID P010131101602, and we folded light curves with the period and the energy resolved pulse profiles of ExpID P010131101602 as the example are presented in Fig. \ref{fig:1}.
\par
Energy-resolved pulse profiles obtained with $Insight$-HXMT for nine energy bands from $2-105$ keV are displayed. The pulse profile of each observation show a similar shape which is mainly characterized by one main peak at $\varPhi _{pulse}\simeq 0-0.5$ and the secondary peak at pulse phase of $0.5-1$ is visible at energies below 10 keV. The pulse profile is rather smooth and does not show the conversion of a single pulse into multiple peaks as in other X-ray pulsars, e.g., Vela X-1 \citep{liu2022variations}. \cite{raichur2010effect} analyzed the data of Cen X-3 that observed by $RXTE$ in 1997 February, the profile shape evolves with energy, comparable to our result. While, Ginga reported an equally strong double-peaked profile at lower energies ($\sim 1-3$ keV, \citealt{nagase1992ginga}). The energy-dependent asymmetric pulse profile caused by two emission regions near the magnetic poles are offset by about 10 degrees \citep{kraus1996analyzing}.

\subsection{Phase-averaged spectrum}
Generally, the X-ray continuum spectra deriving from the accretion column of pulsars is modeled with empirical models. These models are usually composed of the power law component with an exponential cutoff at energies about $12-20$ keV. We modelled the broadband continuum spectra of Cen X-3 with two kinds of commonly used phenomenological models. A power-law with an exponential cutoff (Highecut in the Xspec package) has been widely used in past literatures to model the spectra of accretion pulsars  \citep{staubert2019cyclotron}, in spite of that this model leads to line-like feature in the spectral around the $E_{c}$ \citep{Burderi_2000}. The function of the model is shown below:
\begin{equation}
f(E)=KE^{-\Gamma}\times
\begin{cases}
1&(E\le E_c)\\
\exp\{-(E-E_c)/E_f\}&(E>E_c),
\end{cases}
\end{equation}
where E is the photon energy and K is normalization factor, $\Gamma$ is the photon index of the power law, $E_{f}$ and $E_{C}$ is exponential folding energy and cutoff energy in units of keV respectively. The other form of continuum model is negative and positive power-law with an exponential cutoff (NPEX model, \citealt{makishima1999cyclotron}):
\begin{equation}
f\left( E \right) =K_1\left( E^{-\varGamma _1}+K_2E^{\varGamma _2} \right) \exp \left( -E/E_f \right) .
\end{equation}
In applying this model, $\varGamma _2$ is assigned a value of 2 in order to simulate the Wien hump in the Comptonized radiation. Furthermore, we adopt TBabs (XSPEC) model to describe the observed X-ray spectrum of Cen X-3 is modified by the absorption of X-rays below $\sim 4$ keV by gas and dust composed mainly of hydrogen in the Galaxy. There also exists an emission feature in the spectrum of Cen X-3 at $\sim 6.4$ keV, due to one K-shell electron ejected following neutral (or low ionized) iron in the accretion disk which is photoionized by hard X-rays \citep{10.1093/mnras/stt1947}. We add a Gaussian function to fit the iron emission line.

After the continuum spectral fittings, the significant absorption feature from $\sim 15 - 50$ keV in the residuals (see Fig. \ref{fig:2}). The previous work has reported the cyclotron absorption line around 30 keV \citep{Suchy_2008,10.1093/mnras/staa3477}, we will perform the fitting with adding one absorption component in the model fittings, by using the Gaussian absorption line (gabs) at first: 
\begin{equation}
GABS\left( E \right) =\exp \left. \left\{ -\frac{D_{cyc}}{\sqrt{2\pi}\sigma _{cyc}}\exp \left[ -\frac{1}{2}\left( \frac{E-E_{cyc}}{\sigma _{cyc}} \right) ^2 \right] \right. \right\}.
\end{equation}

We apply above defined continuum models with a Gaussian absorption line and photoelectric absorption to the fitting of the spectra for Cen X-3, the quality of the fit is so good which is indicated by absorption line–like residuals around 40 -- 55 keV. We then added an additional CRSF component to improve the fits. In the case of NPEX continuum model, the $\chi^2$ changed from 738.4 (684 d.o.f.) to 664.6 (681 d.o.f.) for ExpID P010131100111, with a F-test probability of $5\times 10^{-15}$ for the fitting improvement. Only the F-test statistics may not be used to test the significance of an additional spectral feature \citep{protassov2002statistics}. However, the low false alarm probabilities may make the detection of the line stable against even crude mistakes in the computation of the significance \citep{kreykenbohm2004x}. Therefore, we conclude that a statistically acceptable spectral fit can be obtained when two Gaussian absorption lines are included in the spectral model. 
\par
In addition to the gabs model, the asymmetrical profile with the cyclabs model is also widely used to describe the cyclotron absorption line \citep{staubert2019cyclotron},  which is expressed as followed:
\begin{equation}
CYCLABS\left( E \right) =\exp \left\{ -D_f\frac{\left( W_fE/E_{cyc} \right) ^2}{\left( E-E_{cyc} \right) ^2+{W_f}^2} \right\} .
\end{equation}
Thus, we also use the cyclabs model to check the evidence for two CRSFs in Cen X-3 spectrum. The different cyclotron line profile model still confirm the presence of the second line around $\sim 47$ keV (see Table \ref{table:1} with the example for the NPEX continuum model). 

In Table \ref{table:1}, the best fit spectral parameters for two observations with two different continuum models are presented. For a comparison, the NPEX model with two cyclotron absorption line models (gabs and cyclabs) are also shown together. We find that the parameters of cyclotron lines do not change significantly with the choice of continuum models. Ultimately, for ExpID P010131101602, the continuum model of HIGHECUT can be used to obtain a reduced $\chi ^2$ of 0.932 for 673 degree of freedom (d.o.f) and the NPEX model improves the fit further resulting in a $\chi ^2$ of 0.976 for 681 d.o.f. Therefore, the model of X-ray spectrum used in the following part of the paper (including the phase-average spectra in this section and phase-resolved spectra in Section 3.3) is NPEX with the gabs components, the complete model for the spectral fittings in Cen X-3 from $2-105$ keV is described in the following:
\begin{equation}
    I\left( E \right) =N_H*NPEX*CRSF\left( E_1 \right) *CRSF\left( E_2 \right) +FeK\alpha .
\end{equation}

\par
\begin{figure}
    \centering
    \includegraphics[width=.5\textwidth]{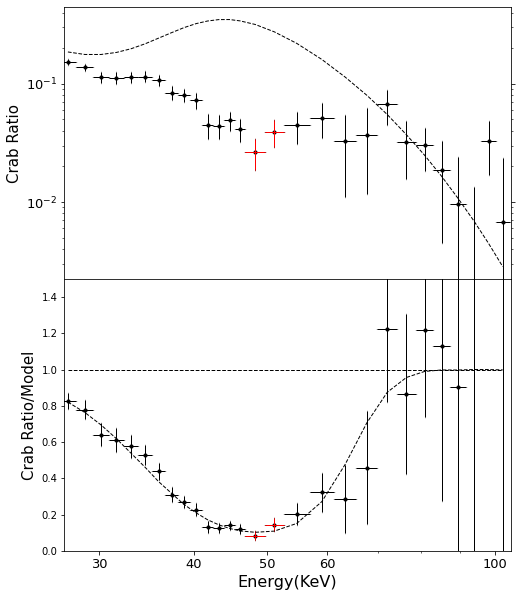}
    \caption{{\bf Top panel}: ratio between Cen X-3 and Crab count rate spectra. The functional form of Cen X-3 divided by the functional form of the Crab spectrum,  i.e., $8.02\times E^{-2.07}$ (see dotted line). {\bf Bottom panel}: ratio between count rate spectra of Cen X-3 and Crab divided by the function of the dotted line in the top panel. The ratio between best fit model with the absorption line at $\sim 47$ keV and model of crab divided by function of the dotted line in the top panel is also shown. }
    \label{fig:3}
\end{figure}

 
In order to check the existence of the CRSF at $\sim 47$ keV in Cen X-3, we examined the ratio between the ExpID P010131101602 and Crab pulsar count rate spectra derived by HE detectors (see Fig. \ref{fig:3}). We used the data of Crab observed by the $Insight$-HXMT in April 2018, when is closest to the observation time of the ExpID P010131101602. The inconsistency between Cen X-3 and Crab's count channel caused by using task $setRebin$ to increase the S/N of source, are shown by the red dots, and we used the Crab's model to replace its real signal. The ratio has the advantage of minimizing the impact of uncertainties in detector response and calibration. Several authors have used the similar method to verify the discovery of the new cyclotron lines (e.g. the absorption feature at 60 keV in Vela X-1 by \citealt{orlandini1997vela}; the second harmonic line of 4U 1626-67 by \citealt{Orlandini_1998}). As seen from the upper panel of Figure \ref{fig:3}, there is a structure at about 47 keV in the Crab ratio spectrum. To more clearly see the effect of cyclotron absorption lines on the continuum, we also use the model of Cen X-3 without an absorption line at $\sim$ 47 keV and other parameters are taken from Table \ref{table:1}. Then, we use this model divided by the model of the Crab Pulsar which we can describe it by fitting in a simple power-law form with an exponent equal to -2.07, we use dotted line in the upper panel of Fig. \ref{fig:3} to describe the ratio between the model of the source and the model of the Crab. In the bottom panel, we show the ratio between the ratio of count rates and the function of dotted line presented in the upper panel of Fig. \ref{fig:3}). In this way we magnify effects due to line features, although we introduce model dependence. Then, we show the ratio between the source's best fit model with two absorption lines and model of crab divided by the function of the dotted line in the top panel as a dotted line. The good agreement between the data and the function indicates that the second absorption line of Cen X-3 is not caused by instrumental effects.

\subsection{Pulse phase-resolved spectroscopy}
We have reported two CRSFs in the average spectral analysis of Cen X-3. The two cyclotron absorption lines would be also detected in the pulse-phase resolved spectra. 
This is acknowledged that the spectrum of an X-ray pulsar varies with the pulse phase, and parameters of cyclotron lines and continuum change with pulse phase. \cite{Molkov_2019} used the ratio of each phase’s spectrum to the pulsed-averaged one to ensure that the cyclotron absorption lines can be detected in different phases, or only appear at certain phase intervals. Thus, the phase-resolved spectroscopy could be considered an effective tool for the research of the magnetic field structure of the emission region. Here we focus on understanding whether the two absorption lines are confirmed in the different pulse phases. We divided the phase for 10 overlapping bins based on the derived pulse periods. For each phase-resolved spectrum, we use the HXMT software to regenerate the new background spectrum and response matrix (also see the methods in \citealt{liu2022variations}).
\begin{figure*}
    \centering
    \includegraphics[width=.46\textwidth,height=12cm]{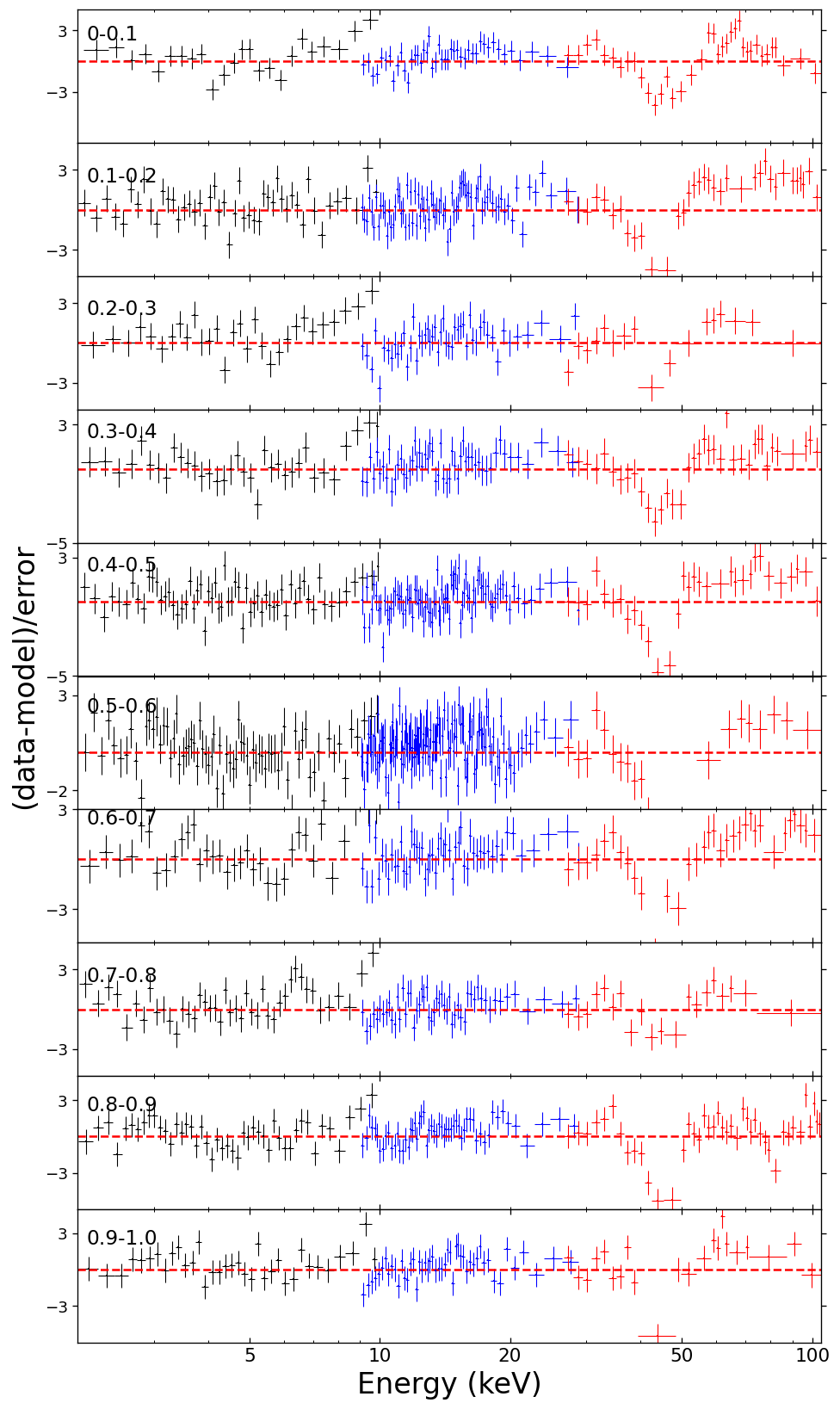}
    \includegraphics[width=.5\textwidth,height=12cm]{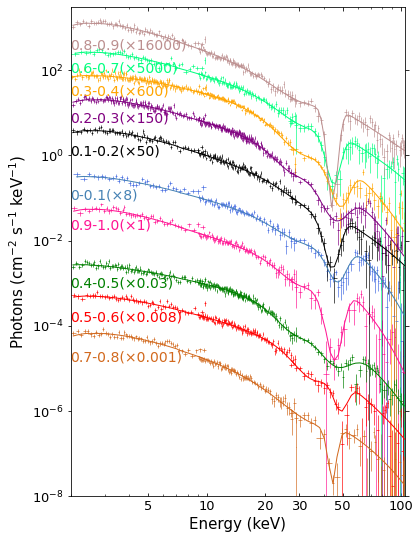}
    \caption{{\bf Left panels:} Residuals of the pulse-phase resolved spectra for ExpID P010131101602 fitting with the NPEX continuum model multiplied by the interstellar absorption and the fundamental line. The absorption features around $\sim 40 - 50$ keV are evidenced in different pulse phases. Phase interval values are given in the each panel. {\bf Right panel:} Phase-resolved spectral profiles in different spin phases, obtained with the NPEX and GABS models for the ExpID P010131101602. Two cyclotron absorption lines at $\sim 28$ keV and $\sim 40-50$ keV are applied in the spectral fittings. }
    \label{fig:4}
\end{figure*}

\begin{figure}
    \centering
    \includegraphics[width=.45\textwidth,height=12cm]{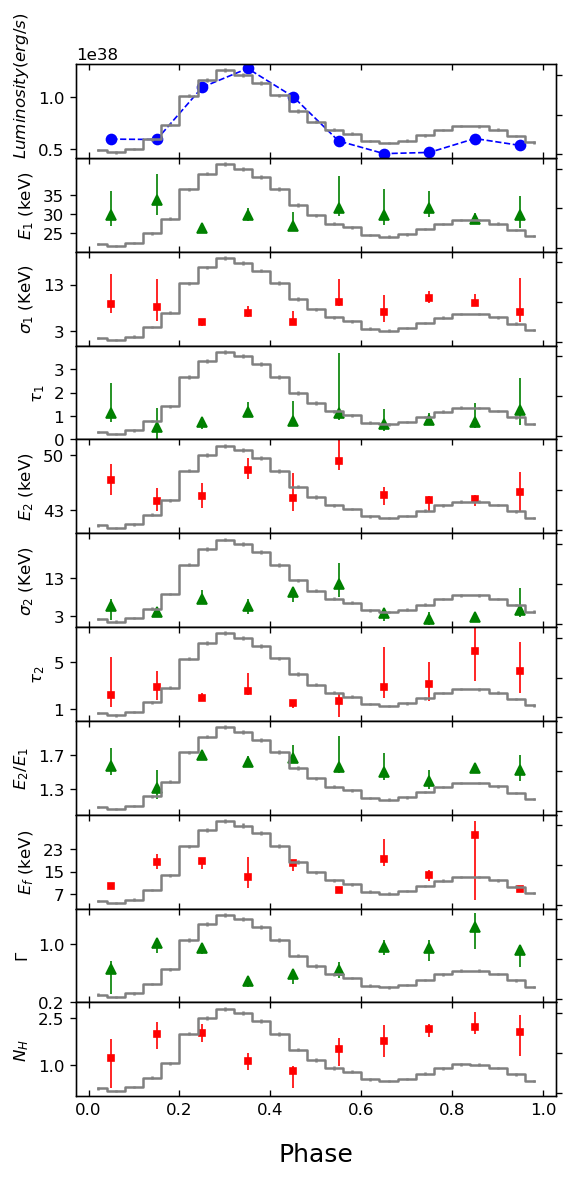}
    \caption{Pulse-phase variations of X-ray luminosity, phase-resolved spectral fitting parameters of the cyclotron absorption lines and the continuum for 10 phase intervals, obtained with the NPEX and GABS models for the ExpID P010131101602. The neutral hydrogen column density $N_{H}$ is in units of $10^{22}$ atoms $cm^{-2}$, $\Gamma$ is the negative photon index in NPEX. The centroid energy, width, depth of the cyclotron absorption lines are represented as $E$, $\sigma$ and $\tau$, respectively, and the gray lines in the panels indicate the pulse profile in the band of 1--10 keV.}
    \label{fig:5}
\end{figure}

\begin{figure}
    \centering
    \includegraphics[width=.48\textwidth,height=8cm]{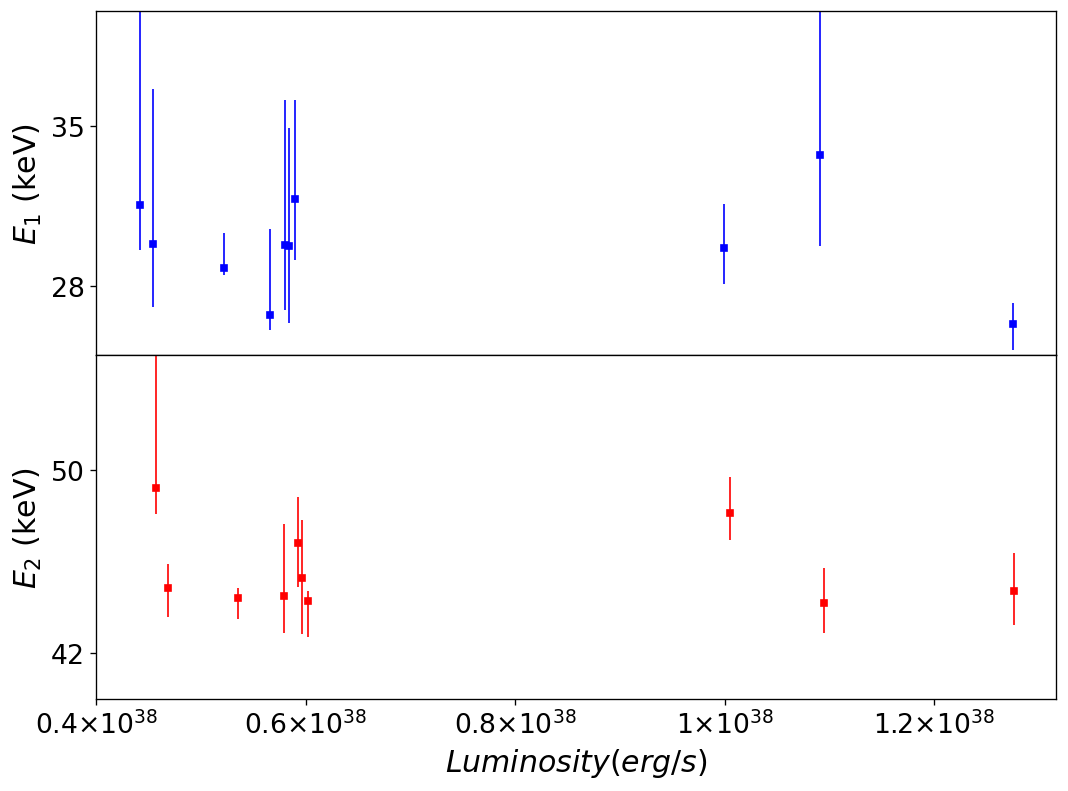}
    \caption{Energies of the fundamental (blue points) and first harmonic (red points) of the cyclotron absorption lines detected in the spectrum of the pulsar Cen X-3 versus its luminosity.}
    \label{fig:6}
\end{figure}
\par
We carried out the phase-resolved spectroscopy of every 0.1 phase interval and at first fitted them with the NPEX continuum model multiplied by the interstellar absorption and the reported fundamental line with the centroid line energy value set between 26 keV and 32 keV by using task $setPars$ in Xspec. Residuals of the fits for all pulse phases are shown in the left panels of Fig. \ref{fig:4}. A significant residual feature can be seen at about 47 keV throughout the pulse phases.
\par
On the next step, the  second absorption line was added to each spectral model. Unlike the phase-averaged analysis, we did not include the iron line since it may be not necessary for the analysis of phase-resolved spectroscopy in X-ray pulsars \citep{molkov2021discovery}. The inclusion of an additional absorption line at $\sim 47$ keV strongly improves the fit quality, i.e., the value of the $\chi ^2$ changes from 729.38 (587 d.o.f.) to 500.87 (477 d.o.f.) at $\varPhi _{pulse}$ of $0.6-0.7$, and at the $\varPhi _{pulse}$ of $0.3-0.4$, from 786.55 (631 d.o.f.) to 635.18 (628 d.o.f.). For phases $0.2-0.3$, we fixed the width of the second cyclotron line to be $\sim 5$ keV obtained from the average value of the other phases, which may be due to the low S/N caused by the insufficient statistics of the phase-resolved spectrum. We also derived the X-ray flux from 2 -- 105 keV for each pulse phase and calculated the luminosity assuming a distance of 6.9 kpc.
\par
Finally, we confirm the existence of two cyclotron absorption features in the phase-resolved spectrum (also see the right panel of Fig. \ref{fig:4}). Variations in the best-fit parameter values of two cyclotron absorption lines over the pulse phase are also presented in Fig. \ref {fig:5}. The fundamental CRSF at $\sim 30$ keV appears to have a similar variation with the X-ray flux profile During the ascent of the pulse profile, the value of cyclotron line is about $30-35$ keV, while between the main and secondary peaks, $E_1\sim 25 - 32$ keV. The fact that the energy of the fundamental cyclotron line reaches the maximum during the off-pulse and ascent phases was also confirmed by \cite{Burderi_2000}. The second cyclotron absorption line evolves throughout the pulse similar to the evolution pattern of $E_1$. $E_2$ increases during the rise of the pulse from $\sim 43$ keV to $\sim 48$ keV for the duration of the main peak. Between the main and secondary peaks, $E_2$ increase to $\sim 50$ keV, and drop back to $\sim 45$ keV at the second peak. The cyclotron widths $\sigma_1$ and $\sigma_2$ vary from $\sim 3-11$ keV, the evolution over the phase follows the behaviour of the cyclotron energy. The absorption depths of the two cyclotron lines vary by almost a stable trend with the pulse phase ($\tau_1\sim 1$ and $\tau_2 \sim 2$), with the deeper lines near the secondary peak. We also plot the ratio of $E_2/E_1$ with different pulse phases in Fig. \ref{fig:5}, the line energy ratio is about 1.4 near $\varPhi _{pulse}\sim 0.2$, and the value stabilizes between 1.6 and 1.7 for other pulse phases. 
\par
The variation of continuum spectral parameters over the pulse phase is also presented in Fig. \ref{fig:5}. The column density $N_{\rm H}$ changes significantly over the pulse phase, it increases from $\sim 1.2\times 10^{22}$ atoms cm$^{-2}$ to $\sim 2.5\times 10^{22}$ atoms cm$^{-2}$ during the rise of the main peak, and decreases to $\sim (0.5-1)\times 10^{22}$ atoms cm$^{-2}$ in the decay of the main peak. And then it changes with the profile of the second peak, with a value of $\sim (1.6-2.2)\times 10^{22}$ atoms cm$^{-2}$. The absolute value of the negative photon exponent in the NPEX which also changes over the pulse phase strongly similar to $N_{\rm H}$, which increases to a value of $\Gamma \sim 1$ during the rise of the main peak, following a decrease to 0.3 during the decay of the main peak, and then jumps back to $\sim 0.8-1$ during the second peak phase.
\par
\section{Conclusion and discussion}
In this work, we investigated the X-ray properties of accreting X-ray pulsar Cen X-3 from $2-105$ keV. In the both phase-average spectral analysis and phase-resolved spectroscopy, we not only confirm the CRSF at $\sim 28$ keV, in particular, we have detected one additional feature at $\sim 47$ keV with the high significance levels, thus making Cen X-3 be one of the X-ray pulsars with more than one cyclotron absorption line. The existence of two CRSFs in Cen X-3 is not dependent on the continuum spectral models (e.g., HighECUT and NPEX) and spectral line profile models (gabs and cyclabs). During the observations with two cyclotron absorption lines reported, Cen X-3 had the mean X-ray luminosity of $\sim 10^{38}$ erg s$^{-1}$ from 2 -- 105 keV assuming a distance of $\sim$ 6.9 kpc. In the phase-resolved spectrum results in Fig. \ref{fig:5}. The energies of both the fundamental and harmonic cyclotron lines evolve throughout the pulse phase, and similar to the shape of the pulse profile. Interestingly, the ratio of two line centroid energies does not change significantly and stabilizes in a narrow value range of $1.6-1.7$ over the pulse phase.
\par
The observed X-ray luminosity varies over the pulse phase from $\sim (4-13)\times 10^{37}$ erg s$^{-1}$. In addition, the centroid line energies of the cyclotron absorption features versus the pulse phase is also derived. In Fig. \ref{fig:6}, we presented energies of both the fundamental (blue point) and first harmonic (red point) of the cyclotron absorption lines detected in the spectrum of the pulsar Cen X-3 versus its luminosity, with the first harmonic line energy ranging from $\sim 43-50 $ keV and the energy of fundamental line in the range of $\sim 25-32$ keV. The centroid line energies have no significant correlation with luminosity, and still show the possible trend of lower line energies at higher luminosity. This relation may connect to a critical luminosity which divides two accretion dynamic regions containing the super-critical and sub-critical cases in accreting pulsars \citep{becker2012spectral}. When $L_{X}>L_{\rm crit}$, plasma is decelerated by the radiation-dominated shock. Below the shock altitude, the photons which break down the balance between downward advection and upward diffusion of radiation are trapped by advection, although they eventually diffuse the column walls to form the observed X-ray spectrum. The emission height varies in proportion to X-ray luminosity in this case which results in an anti-correlation between $E_{\rm cyc}-L_X$. While in the subcritical case, the gas still makes its initial deceleration through a radiation-dominated shock, the final stopping occurs via direct Coulomb interactions caused by accreting ions lose energy to electrons close to the surface of the accretion column. Then the excited electrons return to the ground state via radiative dipole transition so that a significant fraction of these photons will escape through the walls of the column \citep{nelson1993nonthermal}, which makes a positive relation between $E_{\rm cyc}-L_X$. For the case of Cen X-3, we assume that neutron star mass and radius values $M_{*} = 1.21 M_{\odot}$ \citep{ash1999mass}  and $R_{*}$ = 10 km, and $\varLambda$ = 0.1 and $w = 1$ based on the theoretical considerations \citep{becker2012spectral}, $L_{\rm crit}\sim 3.5 \times 10^{37} erg \cdot sec^{-1}$. Thus Cen X-3 is located in a super-critical regime during our observations, and generally an anti-correlation of $E_{\rm cyc}-L_X$ is expected. For the phase-resolved line energies versus the corresponding luminosity (Fig. \ref{fig:6}), the relation is not obvious yet. At present the $E_{\rm cyc}-L_X$ relations are generally reported in the average spectrum of X-ray pulsars \citep{staubert2019cyclotron}, thus much more observational data would help to check the relation in Cen X-3 (work in preparation with the Insight-HXMT data from 2018-2021).

\par
In spectral analysis of Insight-HXMT data, we perform the fit with the analytical models (e.g., NPEX, HighECut), deriving a photon index of $\sim 0.7-1.2$, and the cutoff energy of $\sim 13$ keV with a column density of $\sim (1.8-2.3)\times 10^{22}$ cm$^{-2}$. The observed spectral properties could be also compared with the physical model proposed by \cite{2006Thermal} which calculates emitted spectra from an X-ray pulsar accretion column due to the bulk and thermal Comptonization of bremsstrahlung, cyclotron, and blackbody seed photons. They also modeled and fitted the spectrum of Cen X-3 taking a distance of 8 kpc (6.9 kpc in our case), using a hydrogen column density of $N_H= 2\times10^{22} \rm cm^{-2}$ in order to reproduce the observed low-energy turnover. The cut-off energy could be related to the electron temperature ($T_e\sim 3\times 10^7$ K) of the Comptonizing plasma, and the relatively flat shape at lower energies with $\Gamma <2$ can be ascribed to the contribution of thermal Comptonization. Our measurement on the fundamental cyclotron line energy around 28 keV is consistent with the previous studies and model spectrum of \cite{2006Thermal}, while the harmonic line around 45 -- 48 keV is not included in their model.  

\par
The straightforward interpretation for two cyclotron absorption features is that the absorption line located in $\sim 47$ keV is the first harmonic of the identified 28 keV line as the fundamental CRSF. However, we find that the line energies of two cyclotron lines do not follow the relation $E_2 = 2E_1$, while $E_2/E_1\sim 1.6-1.7$ for the pulse phase-resolved spectra, and $\sim 1.7-1.8$ also for the average spectral cases (see Table \ref{table:1}). A few X-ray pulsars show a similar phenomenon, e.g., MAXI J1409-619 \citep{orlandini2012bepposax}, GX 301-2 \citep{ding2021timing}. This disagreement with the line energy scale relation can be explained by that the first harmonic line is produced by pure absorption \citep{nishimura2003influence}, while for harmonic resonance scattering features, other effects such as multiple scattering and photon spawning also take place \citep{schonherr2007model}. It has been proposed that the absorption features are not due to one line, maybe form independently in different regions \citep{furst2018multiple,ding2021timing}, then the two absorption lines are expected to form at different heights above the surface of the neutron star, the harmonic line situated at the surface of the NS, while the fundamental line at the height ($h\sim 1.5-2$ km based on the ratio of $E_2/E_1$ for Cen X-3) of low the magnetic field strength. Another explanation is to consider by assuming that the center of the dipole magnetic field is offset relative to the center of the neutron star if we assume two absorption lines originate from the different poles of the neutron star \citep{rodes2009first}.

\section*{Acknowledgements}
We are grateful to the referee for the useful comments to improve the manuscript. This work is supported by the National Key Research and Development Program of China (Grants No. 2021YFA0718500, 2021YFA0718503), the NSFC (12133007, U1838103). This work made use of data from the \textit{Insight}-HXMT mission, a project funded by China National Space Administration (CNSA) and the Chinese Academy of Sciences (CAS).
\section*{Data Availability}
Data that were used in this paper are from Institute of High Energy Physics Chinese Academy of Sciences(IHEP-CAS) and are publicly available for download from the \textit{Insight}-HXMT website.
To process and fit the spectrum and obtain folded light curves, this research has made use of XRONOS and FTOOLS provided by NASA.
\bibliographystyle{mnras}
\bibliography{example}




\bsp	
\label{lastpage}
\end{sloppypar}
\end{document}